# Precise Control of Band Filling in Na$_x$CoO$_2$

Daisuke YOSHIZUMI, Yuji MURAOKA**, Yoshihiko OKAMOTO, Yoko KIUCHI,
Jun-Ichi YAMAURA, Masahito MOCHIZUKI[1, ***], Masao OGATA[2] and Zenji HIROI*

*Institute for Solid State Physics, University of Tokyo, Kashiwa, Chiba 277-8581*
[1]*RIKEN, Hirosawa, Wako, Saitama 351-0198*
[2]*Department of Physics, University of Tokyo, Hongo, Bunkyo-ku, Tokyo 113-0033*

Electronic properties of the sodium cobaltate Na$_x$CoO$_2$ are systematically studied through a precise control of band filling. Resistivity, magnetic susceptibility and specific heat measurements are carried out on a series of high-quality polycrystalline samples prepared at 200°C with Na content in a wide range of $0.35 \leq x \leq 0.70$. It is found that dramatic changes in electronic properties take place at a critical Na concentration $x^*$ that lies between 0.58 and 0.59, which separates a Pauli paramagnetic and a Curie-Weiss metals. It is suggested that at $x^*$ the Fermi level touches the bottom of the $a_{1g}$ band at the Γ point, leading to a crucial change in the density of states across $x^*$ and the emergence of a small electron pocket around the Γ point for $x > x^*$.

*E-mail: hiroi@issp.u-tokyo.ac.jp
**Present address: *The Graduate School of Natural Science and Technology, Okayama University, 1-1, Naka-3, Tsushima, Okayama 700-8530*
***Present address: *Tokura Multiferroics Project, ERATO, Japan Science and Technology Agency (JST)*

Layered sodium cobaltate Na$_x$CoO$_2$[1] provides us with a fascinating playground to study the physics of strongly correlated electrons on the frustrated triangular lattice in a wide range of band filling. Recent two important discoveries have accelerated the research of Na$_x$CoO$_2$: one is an unusually large thermoelectric power for $x \sim 2/3$[2] and the other is superconductivity below 4.5 K in a hydrated compound with $x \sim 1/3$.[3] Particularly, a lot of studies have focused on the hydrated superconductor to clarify the mechanism of the superconductivity and to understand underlying electronic states realized in the triangular lattice. In spite of extensive research, however, there still remain some controversial issues possibly resulting from difficulty in the chemistry of the compounds that put obstacles in the way of obtaining high-quality samples.[4, 5]

The electronic phase diagram of Na$_x$CoO$_2$ as a function of $x$ has been proposed by several groups. Foo *et al.* gave a typical one where a charge-ordered magnetic insulator exists at $x = 0.5$ with a "paramagnetic (PM) metal" on the left ($x < 0.5$) and a "Curie-Weiss (CW) metal" on the right ($x > 0.5$).[6] In contrast, Yokoi *et al.* found that the boundary between the two metals is located approximately at $x = 0.6$.[7] The reason for this discrepancy is not known. Moreover, the origin of the change in metallic character has not yet been interpreted in a clear manner. On the charge-ordered insulator at $x = 0.5$, it is known that the ordering of Na ions, which is already present at high temperature, triggers a magnetic as well as a metal-insulator transitions at low temperatures.[6-8]

The electronic states of Na$_x$CoO$_2$ near the Fermi level come from the Co $d$ derived $t_{2g}$ state that splits into an $a_{1g}$ singlet and an $e'_g$ doublet. These bands are filled progressively with $x$, because one Na$^+$ ion donates one electron to the CoO$_2$ layer. According to band structure calculations,[9-14] the $a_{1g}$ band always crosses the Fermi level, irrespective of band filling, which gives a large circular (cylindrical) Fermi surface (FS) around the Γ point in the Brillouin zone. In addition, two important features on the FS are found in the local-density approximation calculations: one is a set of small hole pockets originating from the $e'_g$ band near the K points, which are expected to appear at low fillings such as $x = 0.3$. The other is a small concentric electron FS around Γ coming from a dip in the band energy of the $a_{1g}$ state, which may appear at high fillings such as $x = 0.7$. However, the absence of the latter was suggested for any doping levels by more sophisticated band structure calculations incorporating spin polarizations or the on-site Coulomb interaction $U$.[11] Experimentally, angle-resolved photoemission spectroscopy (ARPES) study on Na$_{0.7}$CoO$_2$ observed only a large circular FS around Γ and failed to detect the other FSs.[15] Recently, Mochizuki and Ogata pointed out in their tight binding model that band dispersions and FS topology change sensitively with the thickness of the CoO$_2$ layer:[16, 17] the hole pockets near K would appear for a thinner CoO$_2$ layer, while the small electron pocket around Γ would be present for a thicker CoO$_2$ layer. Thus, the current status is miles away from a complete understanding of the basic electronic structures of Na$_x$CoO$_2$.

In this letter, we study systematically the transport and thermodynamic properties of Na$_x$CoO$_2$ using a series of polycrystalline samples. The key point of the present study is a novel method used to prepare samples: most of samples studied so far were obtained by a soft-chemistry method at ambient temperature,[3] which might cause an inhomogeneous distribution of Na ions or otherwise lead to a strong tendency for Na ordering at certain fractional compositions such as $x = 0.5$. Particularly, the Na ordering strongly influences the electronic state of the $(CoO_2)^{x-}$ layer and may mask the intrinsic features. We adopted alternatively a solid-state reaction at higher temperatures starting from two end members of $Na_{0.35}CoO_2$ and $Na_{0.70}CoO_2$ and succeeded in controlling $x$ on a scale of 0.01. Moreover, since the high-temperature reaction favoured disordering of Na ions, more intrinsic properties of the $(CoO_2)^{x-}$ layer have been clarified.

We prepared a series of polycrystalline samples of $Na_xCoO_2$ by a solid-state reaction of $Na_{0.35}CoO_2$ and $Na_{0.70}CoO_2$. First, powders of $Na_{0.70}CoO_2$ were synthesized by a reaction of stoichiometric amounts of $Na_2CO_3$ and $Co_3O_4$ in air at 860°C. Sodium deintercalation was then carried out in a 1.0 M $Br_2$ solution in acetonitrile to obtain powders of $Na_{0.35}CoO_2$. The Na content was determined by the inductively coupled plasma-atomic emission spectroscopy (ICP) method. A dozen of samples with intermediate compositions were prepared by reacting the two powders of $Na_{0.35}CoO_2$ and $Na_{0.70}CoO_2$ in an appropriate ratio in a sealed quartz tube at 200°C for 24 hours, followed by slow cooling to room temperature. The Na content of the products was also examined by the ICP analysis. The phase purity was confirmed by means of powder x-ray diffraction. More detail on the sample preparation and characterization will be reported elsewhere.[18] Resistivity and specific heat were measured in a Quantum Design PPMS system, and magnetic susceptibility measurements were performed in a Quantum Design MPMS system.

Figure 1 shows the systematic variation of resistivity $\rho$ with $x$. For clarity, each set of data is divided by the value at 250 K. The $\rho$ of $x = 0.50$ exhibits a smooth, semiconductor-like increase at low temperature with no anomalies for a metal-insulator transition, in contrast to the previous report where a soft-chemically prepared sample of $x = 0.50$ exhibits a sharp rise in $\rho$ below 53 K.[6, 7] We examined our sample by electron diffraction and found that a superstructure coming from Na ordering was almost absent.[18] We think that this semiconducting behavior is due to a partial ordering of Na ions, because even at room temperature the Na ions tend to align.[18] As $x$ increases from 0.50, such a semiconducting increase in $\rho$ is progressively suppressed, and finally a metallic $T$ dependence appears above 0.55. Nevertheless, a small upturn is observed below 20 K, as shown in the inset to Fig. 1, which may be ascribed to weak localization due to a random electrostatic potential from disordered Na distributions.

Compared with the $T$ dependence in $\rho$ for $x \leq 0.58$, which is always concave-downward, those of $x \geq 0.59$ are apparently dissimilar; almost $T$ linear for $x = 0.59$ and concave-upward for $x = 0.66$ and 0.70. The last one shows a steep decrease below 30 K, which is similar as reported previously for $x = 0.75$.[6] Therefore, the $T$ dependence of $\rho$ changes significantly across $x = 0.58 \sim 0.59$.

Another related change is found in the evolution of magnetic susceptibility $\chi$, as shown in Fig. 2. The $\chi$ for $x \leq 0.58$ shown in the left panel is relatively small in magnitude and weakly $T$ dependent, while, in distinct contrast, the $\chi$ for $x \geq 0.59$ is large in magnitude and shows a characteristic CW divergence that is progressively enhanced with increasing $x$. Hence, a substantial change in magnetism must occur across a critical Na concentration $x^*$ between 0.58 and 0.59, accurately corresponding to the above results from resistivity.

Looking in more detail, the $\chi$ of $x = 0.35$ decreases gradually with decreasing $T$ from 300 K, exhibits a minimum at 120 K and then shows a CW divergence at low temperature. This low-temperature CW component may be due to a minor portion of spins that come from defects or impurities and are noninteracting with majority spins. The positive slope above 120 K suggests the existence of a peak in $\chi$ at a higher temperature. In fact, the 0.50 and 0.52 samples seem to show a broad maximum near 300 K. Moreover, a broad peak is clearly observed at around 150 K for $x = 0.56$. The peak is possibly further shifted to a lower $T$ for $x = 0.58$ and becomes obscure. This broad peak indicates the development of antiferromagnetic (AF) short-range order on cooling. Then, its systematic shift to lower temperatures implies that the characteristic energy of AF spin fluctuations or effective superexchange $J_{eff}$ decreases with increasing $x$. Since the density of spins $(1-x)$ decreases as $x$ increases, we expect that $J_{eff}$ between nearest-neighbour spins also decreases, consistent with the observed behavior in $\chi$. For $x \geq 0.59$, on the other hand, the Weiss temperature deduced from fitting the data to the CW law $\chi = C/(T - \Theta)$ is always negative in sign (AF) and gradually increases from -156 K ($x = 0.59$) to -99 K ($x = 0.70$). This fact indicates that AF fluctuations are suppressed or additional ferromagnetic interactions are enhanced with increasing $x$.

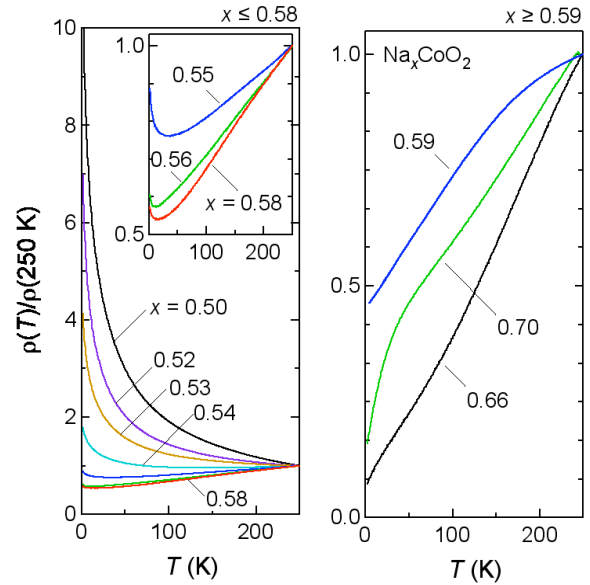

Fig. 1. Evolution of the temperature dependence of resistivity $\rho$ normalized to the value at 250 K for $0.50 \leq x \leq 0.58$ (left) and $0.59 \leq x \leq 0.70$ (right).

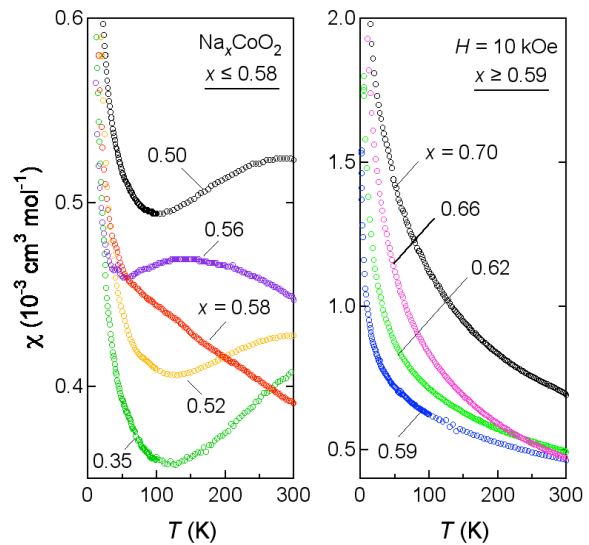

Fig. 2. Magnetic susceptibility $\chi$ measured on heating in a magnetic field of 10 kOe for $0.35 \leq x \leq 0.58$ (left) and $0.59 \leq x \leq 0.70$ (right). Note the difference in the vertical scale.

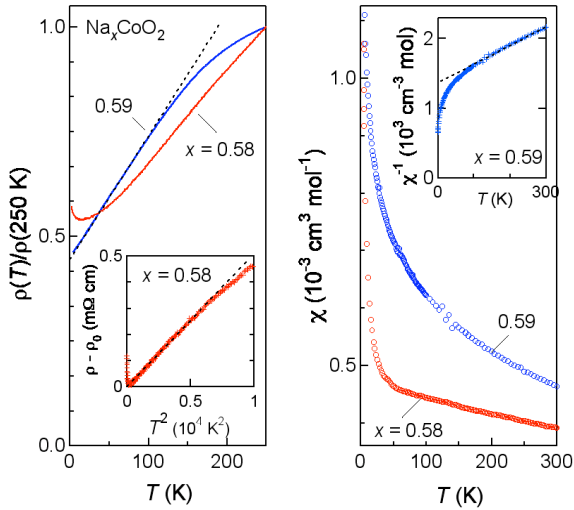

Fig. 3. Comparisons of resistivity (left) and magnetic susceptibility (right) between $x = 0.58$ and 0.59. Inset in the left panel shows a $\rho - \rho_0$ versus $T^2$ plot for 0.58, where $\rho_0$ is the residual resistivity and is 1.83 mΩ cm. Inset in the right panel shows the $T$ dependence of inverse $\chi$ for 0.59.

In order to demonstrate more explicitly the dramatic changes across $x^*$, the $\rho$ and $\chi$ data of $x = 0.58$ and 0.59 are compared in Fig. 3. The $\rho$ of 0.58 is proportional to $T^2$ in a wide $T$ range below 70 K, as shown in the inset, provided that the low-temperature upturn is ignored. Such $T^2$ behavior is what one expects generally for a strongly correlated electron system. In contrast, the $T$ dependence in $\rho$ of 0.59 is rather unusual, showing an almost linear variation in a similarly wide $T$ range below 80 K. Hence, there must be a substantial difference in the scattering mechanism of carriers between the two samples.

On the other hand, it is apparent from Fig. 3 that $\chi$ is enhanced enormously from 0.58 to 0.59. Moreover, the $T$ dependence above 50 K changes from linear to CW-like. This $T$ linear behavior for 0.58 may be accidental as a few contributions of different origins coexist. The unusual $T$-linear dependence in $\rho$ for 0.59 must be related to the appearance of the CW component in $\chi$: additional magnetic scattering can be the source.

To evaluate the change in the $T$ dependence of $\chi$, the slope at 100 K is plotted in Fig. 4, following the previous analysis done by Yokoi et al.[7] The slope is close to zero or takes small negative values for $x \leq 0.58$, while it decreases almost discontinuously to a relatively large negative value at 0.59, followed by a further decrease with increasing $x$.

Another experimental evidence to support the existence of a critical $x$ value has been obtained from specific heat measurements. The $x$ dependence of the Sommerfeld coefficient $\gamma$ determined from the intercept of the $C/T$ versus $T^2$ plot is shown in Fig. 4. Obviously, there is a change in $\gamma$ at $x^*$: $\gamma$ increases slightly with $x$ for $x \leq 0.58$ and suddenly rises at $x^*$ by 17 %, from 16.8 mJ K$^{-2}$ mol$^{-1}$ for 0.58 to 19.7 mJ K$^{-2}$ mol$^{-1}$ for 0.59. Then, it increases rapidly to be saturated at 31.5 mJ K$^{-2}$ mol$^{-1}$ for $x = 0.70$. Note that the $\gamma$ of 0.50 is finite because of the metallic nature of our sample in the absence of charge order. Since $\gamma$ is proportional to the density of states (DOS), we conclude that the DOS at the Fermi level suddenly increases above $x^*$.

All the above results indicate that the electronic structure of Na$_x$CoO$_2$ does not change smoothly with $x$, but there is a well-defined boundary at $x^*$. It is found that the critical concentration lies between 0.58 and 0.59, not at 0.5 as reported by Foo et al.,[6] but close to 0.60 as reported by

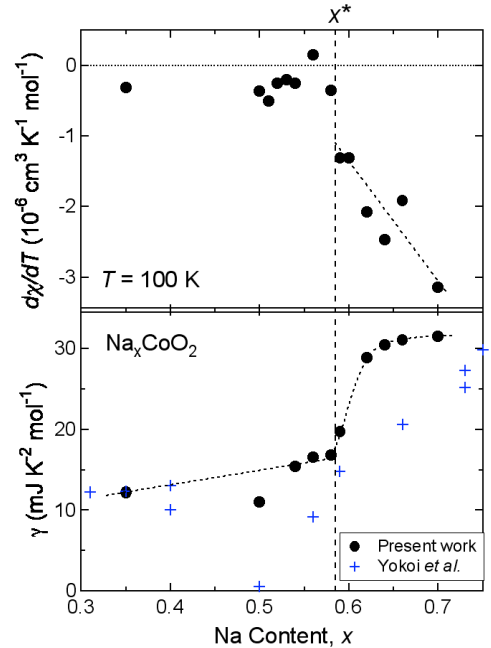

Fig. 4. $x$ dependences of the slope in $\chi$ at 100 K (top) and the Sommerfeld coefficient $\gamma$ (bottom). Crosses in the bottom panel represent the data reported by Yokoi et al.[7] Dotted lines serve as guides to the eye.

Yokoi et al.[7] The sharp boundary breaks the phase diagram into two regions: a Pauli paramagnetic metal with a relatively small DOS for $x < x^*$ and a CW metal with an enhanced DOS for $x > x^*$ (Fig. 5). A main reason why we observed such sharp changes in properties at $x^*$ in the present study may be ascribed to the high quality of our samples prepared at high temperature as well as the reduced influence of the Na ordering.

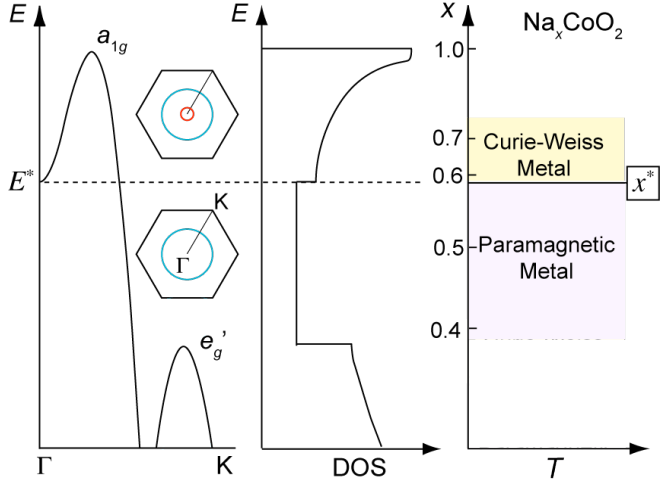

Fig. 5. Schematic representation of the band structure of $Na_xCoO_2$. Band dispersions along the Γ-K line (left), an expected profile of density of states (middle) and an $x$-$T$ phase diagram (right) are depicted. The critical Na content $x^*$ in the phase diagram corresponds to a band filling with the Fermi energy equal to $E^*$, as shown by a broken line, where the Fermi level touches the bottom of the $a_{1g}$ band at the Γ point. Hexagons represent the Brillouin zone with the Γ point at the center and the K points at the corners. A small electron Fermi surface appears for the Fermi energy above $E^*$ in addition to a large Fermi surface around Γ. In the phase diagram, a Curie-Weiss metal exists above $x^*$, while a Pauli paramagnetic metal below $x^*$. A charge-ordered insulator reported at $x = 0.5$ is excluded in this phase diagram, because it is extrinsic, coming from the Na ordering.

On the basis of these experimental lines of evidence, we are now ready to consider what is going on with $x$ in $Na_xCoO_2$ in terms of band structures. As $x$ decides filling in the $t_{2g}$ band, it is reasonable to assume that the band structure changes at a certain filling in the rigid band picture. According to band structure calculations,[9-13] there is a dip around the Γ point in the $a_{1g}$ band, though it has not yet been detected experimentally.[15] Then, it is likely that at $x^*$ the Fermi level with energy $E^*$ touches the bottom of the $a_{1g}$ band exactly at the Γ point, as schematically depicted in Fig. 5. An approximate profile of DOS expected from the band structure is also illustrated in Fig. 5, which is constant below $E^*$ due to the two dimensionality and exhibits a discontinuous jump at $E^*$, followed by a further increase above $E^*$. This profile of DOS is qualitatively in good agreement with the observed $x$ dependence of $\gamma$ shown in Fig. 4. A minor difference between them may come from the actual three dimensionality of the band structure that splits the $a_{1g}$ band into two.[9, 12] On the other hand, the top of the $e'_g$ band must be below the Fermi level even at the lowest filling, because no enhancement in $\gamma$ is observed for $0.35 \le x < x^*$ in Fig. 4.

Furthermore, one expects a distinct change in the topology of the FS at $x^*$: an additional small electron pocket should appear for $x > x^*$. Because of this small electron pocket, strong ferromagnetic correlations are expected for $x > x^*$, which has been predicted theoretically[10, 13, 19] and evidenced by neutron diffraction[20, 21] and other experiments.[22-24] The observed change from a PM metal to a CW metal is obviously attributed to the emergence of this small electron FS.

It seems difficult to estimate the accurate value of $x^*$ from band structure calculations, because the band structure is highly sensitivity to the crystal structure or the effect of electron correlations.[9-14] However, Korshunov et al. estimated $x^*$ ($x_m$ in their report) to be 0.56 based on a tight-binding fit to an LAPW calculation, which is close to our value.[13, 14] They pointed out that taking account of electron correlations would push it up to 0.68,[13, 14] which suggests that electron correlations may not be so important in $Na_xCoO_2$.

Concerning superconductivity found in the hydrated compound, Sakurai et al. found that superconductivity appears at two specific Co valences of +3.48 and +3.40.[4, 5] Interestingly, the latter corresponds to $x = 0.60$ in $Na_xCoO_2$, just above our $x^*$. Kuroki and his coworkers pointed out theoretically that the presence of the two concentric FSs around the Γ point leads to an enhanced spin fluctuation and thus gives rise to an extended s-wave superconductivity in the hydrated compound.[25] We think that superconductivity would show up at just above $x^*$ even in nonhydrated $Na_xCoO_2$, if the influence of disordered Na ions is appropriately taken away. The other Co valence of +3.48 corresponds to $x = 0.52$ in $Na_xCoO_2$, which is a simple PM metal. However, a neutron diffraction experiment found that hydration squashes the $CoO_2$ layer.[26] It is theoretically predicted that this structural change with hydration pushes the $e'_g$ band above the Fermi level, leading to an enhancement in DOS and thus the occurrence of a spin-triplet superconductivity.[16, 17] Therefore, the two superconducting states may be associated with the two corresponding FSs with enhanced DOS.

In summary, we have studied the electronic properties of $Na_xCoO_2$ with varying $x$, using a series of high quality samples prepared at high temperature. Dramatic changes in various quantities and thus in the electronic structure are found at a critical Na concentration $x^*$ between 0.58 and 0.59. This provides strong evidence of the presence of an electron pocket around the Γ point for high band fillings of $x > x^*$. The intrinsic phase diagram of the $(CoO_2)^{x-}$ layer is rather simple, as depicted in Fig. 5, in the absence of an electrostatic potential superimposed from the Na layers above and below.

**Acknowledgment**

We thank M. Ichihara for his help in electron microscopy observations.

1) C. Fouassier *et al.*: J. Solid State Chem. **6** (1973) 532.
2) I. Terasaki, Y. Sasago and K. Uchinokura: Phys. Rev. B **56** (1997) R12685.
3) K. Takada *et al.*: Nature **422** (2003) 53.
4) H. Sakurai *et al.*: J. Phys. Soc. Jpn. **74** (2005) 2909.
5) H. Sakurai *et al.*: Phys. Rev. B **74** (2006) 092502.
6) M. L. Foo *et al.*: Phys. Rev. Lett. **92** (2004) 247001.
7) M. Yokoi *et al.*: J. Phys. Soc. Jpn. **74** (2005) 3046.
8) G. Gaparovi *et al.*: Phys. Rev. Lett. **96** (2006) 046403.
9) D. J. Singh: Phys. Rev. B **61** (2000) 13397.
10) D. J. Singh: Phys. Rev. B **68** (2003) 020503(R).
11) P. Zhang *et al.*: Phys. Rev. Lett. **93** (2004) 236402.
12) M. D. Johannes *et al.*: Europhys. Lett. **68** (2004) 433.
13) M. M. Korshunov *et al.*: JETP Letters **84** (2007) 650.
14) M. M. Korshunov *et al.*: Phys. Rev. B **75** (2007) 94511.
15) M. Z. Hasan *et al.*: Phys. Rev. Lett. **92** (2004) 246402.
16) M. Mochizuki and M. Ogata: J. Phys. Soc. Jpn. **75** (2006) 113703.
17) M. Mochizuki and M. Ogata: J. Phys. Soc. Jpn. **76** (2007) 013704.
18) D. Yoshizumi *et al.*: in preparation.
19) K. Kuroki *et al.*: Phys. Rev. Lett. **98** (2007) 136401.
20) S. P. Bayrakci *et al.*: Phys. Rev. B **69** (2004) 100410(R).
21) A. T. Boothroyd *et al.*: Phys. Rev. Lett. **92** (2004) 197201.
22) K. Ishida *et al.*: J. Phys. Soc. Jpn. **72** (2003) 3041.
23) I. R. Mukhamedshin *et al.*: Phys. Rev. Lett. **94** (2005) 247602.
24) Y. Ihara *et al.*: J. Phys. Soc. Jpn. **75** (2006) 124714.
25) K. Kuroki *et al.*: Phys. Rev. B **73** (2006) 184503.
26) J. W. Lynn *et al.*: Phys. Rev. B **68** (2003) 214516.